\documentclass[a4paper,10pt]{article}
\usepackage[geometry]{ifsym}
\usepackage[english]{babel}
\usepackage{verbatim}
\usepackage{hyperref}
\usepackage{color}
\usepackage{soul}
\usepackage{mathrsfs}
\usepackage{amsfonts,amsmath,latexsym}
\usepackage[T1]{fontenc}
\usepackage{hyperref}
\usepackage{graphicx}
\usepackage{enumerate}
\usepackage{epsf}
\usepackage{natbib}
\usepackage{amsthm}
\usepackage{color}
\usepackage{amssymb}
\usepackage{fancyhdr} 
\usepackage{palatino}

\renewcommand{\theequation}{\arabic{section}.\arabic{equation}}

\def\R{\mathbb R}
\def\N{\mathbb N}

\author{
{\sc}\ {\sc Jia Liu}
\thanks{Corresponding author, Department of Mathematics and Statistics,  University of Jyv\"askyl\"a, P.O.Box (MaD) FI-40014  Finland e-mail:{\tt  jia.liu@jyu.fi}}
\ {\sc Dario Gasbarra}
\thanks{ Department of Mathematics and Statistics,  University of Helsinki   P.O. Box 68 
FI-00014   Finland e-mail:{ \tt dario.gasbarra@helsinki.fi}}
\ {\sc and}\ {\sc Juha Railavo} 
\thanks{HUS e-mail:{\tt juha.railavo@elisanet.fi}}.
}

\date{December 2014}
\title{Fast Estimation of Diffusion Tensors under Rician noise by the EM algorithm}
\newcommand{\MBFigure}[6]{
$\left. \right.$ \\
\refstepcounter{figure}
\addcontentsline{lof}{figure}{\numberline{\thefigure}{\ignorespaces #5}}
\begin{center}
\begin{minipage}{#1cm}
\centerline{\includegraphics[width=#2cm,angle=#3]{#4}}
\begin{center}
\upshape{F\textsc{ig} \normal
\end{center}
size{\thefigure}. $-$} #5
\end{center}
\label{#6}
\end{minipage}
\end{center}
$\left. \right.$ \\}
 
\begin{document}
\maketitle 

\begin{abstract}
Diffusion tensor imaging (DTI) is widely used to characterize, in vivo, the white matter of the central nerve system (CNS). This biological tissue contains much anatomic, structural and orientational information of fibers in human brain. Spectral data from the displacement distribution of water molecules located in the brain tissue are collected by a magnetic resonance scanner and acquired in the Fourier domain. After the Fourier inversion, the noise distribution is Gaussian in both real and imaginary parts and, as a consequence, the recorded magnitude data are corrupted by Rician noise. 

Statistical estimation of diffusion leads a non-linear regression problem. In this paper, we present a fast computational method for Maximum Likelihood estimation (MLE) of diffusivities under the Rician noise model, based on the Expectation Maximization (EM) algorithm. By using data augmentation, we are able to transform a non-linear regression problem into the Generalized Linear Modeling (GLM) framework, reducing dramatically the computational cost. 
The Fisher-scoring method is used for achieving fast convergence of the tensor parameter. The new method is implemented and applied using both synthetic and real data in a wide range of $b$-amplitudes up to 14000 $s/mm^2$. Higher accuracy and precision of the Rician estimates are achieved compared with other log-normal based methods. In addition, we extend the ML framework to the maximum a posterior (MAP) estimation in DTI under the aforementioned scheme by specifying the priors. 
We will describe 
how close numerically are the estimators of model parameters obtained through ML and MAP estimation.
\end{abstract}



\section{Introduction}
\label{sec:intro}
Diffusion tensor imaging (DTI) 
is a powerful tool to detect, in vivo, the  white matter anatomy and structures of the brain.
The raw MR-data are collected by a magnetic resonance scanner and consist of spectral measurement from the displacement distribution of water molecules constrained into cellular structures. Diffusion anisotropy characterizes the nervous fibers. 

After the Fourier inversion, the  MR-signals are corrupted by a complex Gaussian noise, and consequently, the recorded measurement magnitudes,
referred as diffusion weighted magnetic resonance imaging (DW-MRI) data, will follow the Rician distribution. 
The noise distribution, however, will still stay Gaussian  in both real and imaginary components. 
The simplest method for diffusion tensor estimation (DTE) is based on the linearized log-normal regression model, where the residual variance is assumed to be either constant (Least Squares) 
or depending on the signal amplitude (Weighted Least Squares).  
These Gaussian noise models fail to  fit the high frequency data,  
which carry information about the higher order  diffusion characteristics. 
In the existing literature \cite{rajan2011,veraart2011,andersson2008} on the ML-estimation of diffusion tensors 
under the Rician noise, the maximization algorithm involves repeated  computation of modified
Bessel functions. 
By using data augmentation we are able to replace the Rician likelihood by a Poisson likelihood which is standard in the framework of GLM.

Such simplification reduces dramatically the  computational burden of 
the Fisher-scoring maximization algorithm. This applies also at high $b$-amplitudes, 
where in the low signal regime measurements below a threshold are customarily  coded as zeros. 
In the standard LS or WLS approaches,  zero-measurements are problematic since they cannot be fitted by a log-normal distribution, and simply discarding them 
induces selection bias.
The appropriately modeled noise level provides capability of data correction in further insights, e.g. removing artefacts from the raw data.
 
This paper is structured as follows. Section \ref{sec:theo} describes 
data augmentation and specifies the statistical model for DTE. 
In Section \ref{sec:method} we discuss the implementation of the EM and the Fisher-scoring algorithms in the DTI  context. 
In addition,  we also specify priors for the parameters and  discuss the  computation of the  
Maximum a  Posteriori  Estimator (MAPE) under the same scheme.  Section \ref{sec:res} illustrates the results from both synthetic
and real data. In Section \ref{sec:dis} we conclude with an overview of the methods and the undergoing developments.

\section{GLM for MRI observations}
\label{sec:theo}
\subsection{Rician noise in MRI}
In magnetic resonance imaging (MRI), we usually need to take the noise in the raw MR-acquisitions into account. The complex valued noise $\epsilon$ is composed of two $i.i.d.$ Gaussian random variables with zero mean and variance $\sigma^2$, one for the real and the other one for the imaginary component. After the Fourier inversion, the signal intensity $S\geq 0$ is corrupted by the the complex Gaussian noise, and $Y=|S+ \epsilon|$ will be observed.

Consequently, the observed MR-signal magnitudes follow a Rician distribution resulting in the likelihood function 
\begin{align}\label{(1)}
p_ {S, \sigma^2}( y ) = \frac{ y }{\sigma^2} 
\exp\biggl(  - \frac{ y^2+S^2 }{2\sigma^2} \biggr)
I_0\biggl(  \frac{ y S }{\sigma^2}\biggr) , 
\end{align}
where $ I_\alpha$ is the $\alpha$-order modified Bessel function of first kind. For $\alpha=0$ it
has also the following representation in terms of  Gaussian hypergeometric series \cite{tables}:
\begin{align}\label{special_representation} 
I_0(2 \tau)= {}_0 F_1(1,\tau^2)= \sum_{n=0}^{\infty}  \frac{ \tau^{2n} } {(n!)^2}.
\end{align} 
Let 
$t= S^2/(2\sigma^2)$, then Eq. (\ref{(1)}) gives 
 \begin{align} \label{(2)}
P_{t,\sigma^2}( Y \in dy) = 
\frac y {\sigma^2} \exp\biggl( -t - \frac{ y^2}{ 2\sigma^2} \biggr) I_0 \biggl( \frac{ y }{\sigma }\sqrt{ 2t  } \biggr) dy
\end{align}
with 
$\tau =  y S/(2 \sigma^2 ) = \sqrt{ 2t  } y /(2 \sigma ) $.
\subsection{Data augmentation}
We follow the strategy presented in \cite{Gasbarra} introducing an augmented data $N$ from a Poisson distribution with mean $t>0$. 
The likelihood of the observed data can be transformed from the Rician likelihood Eq. (\ref{(2)}) to a joint augmented density
\begin{align}\label{eq:joint}&
  P_{t,\sigma^2} (N=n, Y^2 \in dy^2) 
 = P_{t,\sigma^2} (N=n, X \in dx)&\\& \notag
 =  P_t( N=n) P_{\sigma^2}(X\in dx| N=n)
=\frac {(tx) ^{n }   }{ (n!)^2  (2\sigma^2)^{n+1} } \exp\biggl(-t -  \frac x {2\sigma^2} \biggr ) dx \; ,
\end{align}
where $X$ is from the conditional distribution Gamma($N+1, 1/(2\sigma^2)$) given $N$.
Eq. (\ref{eq:joint}) provides a transformation from a non-linear regression problem to the GLM framework  
\begin{align}\label{eq:glm} 
f_{\xi,\phi}(z)=c(z,\phi) \exp\biggl(\frac{ z\xi- a(\xi) }{\phi} \biggr) \; 
\end{align}
with $z$ corresponding to the response in general, see \cite{mc_cullagh-nelder} for more details.
\section{Method}
\label{sec:method}
\subsection{DW-MRI and parametrization}\label{subsec:para}
In  DW-MRI, the signal is modeled as the first equality
\begin{align*}
 S( {\bf q} )= S_0 \exp\bigl( - b d({\bf g }) \bigr) = S_0 \exp\bigl(Z\theta  \bigr),
\end{align*}
where the control vector ${\bf q} \in \R^3$ is determined by the sequence of gradient pulses, 
$b= |{\bf q}|^2$,  and ${\bf g}= {\bf q}/| {\bf q}|\in {\mathcal S}^2$ is a vector of unit length.
The  MR-signal decays exponentially with respect to the $b$-amplitude. Depending on the gradient direction ${\bf g}$ the decay is modeled by the  reflection symmetric  diffusivity  function  $d: {\mathcal S}^2 \to \R^+$.

Great efforts have been devoted to modeling the diffusivity, and in general we can have parametrization as the second equality. In the simplest model the diffusivity is expressed by a symmetric and  positive definite rank-2 tensor $D\in \R^{3\times 3}$, giving
\begin{align*}&
\log S( {\bf q} )=  \log S_0 - b  {\bf g}^{\top}  {D}  {\bf g } = \log S_0+ Z\theta  \; ,
\end{align*} 
where 
in the left hand side the  diffusion tensor is parametrized as
\begin{align*} \theta=(\theta_1,\dots, \theta_6)^{\top}:=  
\bigl( D_{xx}, D_{yy} , D_{zz}, D_{xy}, D_{xz}, D_{yz} \bigr)^{\top}
\end{align*}
with a design matrix 
\begin{align*} 
Z= Z ({\bf q})  =  -b\bigl( {\bf g}_x^2  ,   
{\bf g}_y^2 ,   {\bf g}_z^2 , 2 {\bf g}_x    {\bf g}_y , 2 {\bf g}_x   {\bf g}_z , 2 {\bf g}_y  {\bf g}_z       \bigr) \; . 
\end{align*} 
In high angular resolution models (HARDI) (see e.g. \cite{barmpoutis2009}), the diffusivity  is modeled with a totally symmetric cartesian tensor $D$ of order $n\in \N$, as
\begin{align*}
 d({\bf g} ) :=\sum\limits_{\ell_1 =1}^3 \sum\limits_{\ell_2 =1}^3 \cdots \sum\limits_{\ell_{2n} =1}^3  D_{\ell_1, \ell_2, \dots ,\ell_{n}} 
 g_{\ell_1} g_{\ell_2}\cdots g_{\ell_{2n}}  \; .
\end{align*}

 \subsection{EM in MLE}\label{subsec:mle}
In the optimization of the likelihood, we employ the EM (Expectation - Maximization) algorithm, which is one among the iterative methods in the MLE or in the Maximum a Posterior Estimation (MAPE). 
The EM algorithm proceeds in two steps and shortens the computational complexity by using
augmented data. In terms of our case, in the E-step we calculate the expectation of the log likelihood w.r.t the conditional distribution of $N$ given by the observations and other parameters with fixed values. 
In the M-step, we find the ML parameter of $S_0^2$ and $\sigma^2$ by maximizing the augmented log likelihood quantities.
The computational details are listed in Appendix \ref{ape:A}.

The log likelihood from Eq. (\ref{eq:joint}) 
is expressed as
\begin{align}\label{likelihood}&
Q:= \log \bigl(  p_{t,\sigma^2} (N=n, Y ) \bigr)=  c(Y,N)+ N\log(t) - (N+1)\log(\sigma^2) -  t-\frac{Y^2}{2\sigma^2},  
\end{align} 
where $c(Y,N) = N\log(Y^2) - 2\log(N!)- (N+1)\log(2)$ does not depend on $(t,\sigma^2)$ which will be omitted in the M-step. From Section \ref{subsec:para}, we have $t=  S_0^2 \exp(2Z\theta)  / 2\sigma^2$. 

In the EM-iteration, given the current parameter estimates $(\theta^{(k)}, {S_0^2}^{(k)},{\sigma^2}^{(k)})$, we update the conditional expectation of the augmented data by
\begin{align*}
 \langle N \rangle^{(k)} := E_{t^{(k)},{\sigma^2}^{(k)}}\bigl(  N \big\vert Y \bigr ) = \frac{ \tau^{(k)}  \; I_1\bigl( 2 \tau^{(k)} \bigr )  }{ I_0\bigl( 2\tau^{(k)} \bigr) } \; 
 \quad \mbox{ with } \quad \tau^{(k)}= \frac{ Y  S_0^{(k)} \exp(Z \theta^{(k)} ) }{  2 {\sigma^2}^{(k)} }
\end{align*}
In the M-step we update $\sigma^2$ and $S_0^2$ by the recursions 
\begin{align}\label{eq:sigma}
 (\sigma^{(k+1)})^2 =    \biggl(\sum\limits_{i=1}^m \bigl((S_0^{(k)})^2 \exp\bigl( 2 Z_i \theta^{(k)} \bigr)  + Y_i^2 \bigr) \biggr) \bigg/ \biggl(
2 m+4 \sum\limits_{i=1}^m  \langle N_i \rangle^{(k)} \biggr) \; 
\end{align}
and
\begin{align}\label{eq:S0}
   (S_0^{(k+1)})^2 =  2( \sigma^{(k)} )^2 \biggl( \sum\limits_{i=1}^m \langle N_i \rangle^{(k)} \biggr)
   \bigg / \biggl(\sum\limits_{i=1}^m \bigl(\exp\bigl( 2 Z_i \theta^{(k)} \bigr)  \biggr) \; ,
  \end{align}
where $m$ is the number of acquisitions at each voxel.

For the tensor parameter $\theta$, we employ a stabilized Fisher scoring method: given the stabilizing parameter  $\alpha\in [0,1]$, we iterate the recursion
\begin{align}\label{eq:theta}
  \theta \to   \theta + \biggl( (1-\alpha) J(\theta) + \alpha S(\theta)^{\top} S(\theta) \biggr)^{-1}   S(\theta)  , \; 
\end{align}
until convergence to a fixed point \cite{lange}. 
In Eq. \eqref{eq:theta} the score $S(\theta)$ is given by
\begin{align*}
  S(\theta) =  2  \sum\limits_{i=1}^m Z_i \langle N_i \rangle^{(k)}-  \bigl(S_0^{(k)} / \sigma^{(k)}  \bigr)^2\sum\limits_{i=1}^m\exp(2Z_i\theta) Z_i^{\top} \;,
\end{align*}
and the corresponding Fisher information is
\begin{align*}
  J(\theta) = 2 \bigl(S_0^{(k)} / \sigma^{(k)}  \bigr)^2\sum\limits_{i=1}^m\exp(2Z_i\theta) Z_i^{\top}  Z_i   \; .
\end{align*}

The initials of the EM algorithm can be obtained through the least square (LS) from a truncated dataset with the diffusion weighting ranging from $0 \sim 1000 s/mm^2 $ in order to fit the Gaussian model (see \cite{jones2004}, \cite{barber1998}).  To pursue higher quality of the initials, 
we could further apply the weighted least square (WLS) described in \cite{zhu2007}. 
In the Appendix \ref{ape:B} we compare the differences between our EM algorithm and the direct optimization of the Rician likelihood in Eq. \eqref{(1)}, which is commonly used to compute the MLE in DTI.
It should be noted that the EM algorithm is needed because of the latent augmented variables; it does not decrease the marginal likelihood of the data, see Appendix \ref{ape:AA} for the proof.

\subsection{EM in MAPE}
In the Bayesian framework, 
the Maximum a Posterior Estimation (MAPE) aims to obtain the point estimates by maximizing the posterior density. 
The difference between MLE and MAPE in this scenario is in the prior probability $\pi(\xi)$.  Given the data $y$, the normalizing constant in the posterior density $\pi(\xi | y)$ does not depend on the parameter $\xi$. We find the MAPE by maximizing the joint density $ \pi( \xi)p_{\xi}(y)$,
and this is achieved by iterating the EM-recursion with the  penalization $\log \pi(\xi)$
\begin{align} \label{EM-MAP:iteration}
   \xi^{(k+1)} = \arg\max_{\xi \in \Xi} 
   \biggl\{  
   E_{\xi^{(k)} }\bigl( 
   \log p_{\xi}( z,y )  
   \big\vert y \bigr) + \log\pi(\xi) \biggr\}\; 
\end{align}
until convergence to a fixed point. The log-prior penalization term has a regularizing effect,
which vanishes asymptotically as the sample size grows \cite{andersson2008}.  

In DTE, we can assign conjugate priors in light of Section \ref{subsec:mle} for $\sigma^2$ and $S_0^2$. Since we have little knowledge of the tensor parameter $\theta$, we may choose non-informative priors which are either scale- or shift-invariant
\cite{jaynes}. A simple Bayesian hierarchical model is obtained after the following choices:

\begin{itemize}
\item $\sigma^2$ has 
scale invariant improper prior with density  $\pi(\sigma^2) \propto  1 / \sigma^2 $,
\item
 $S_0^2 \sim \mbox{Gamma}(c_1, c_2) $, where $c_1, c_2$ are very small.
\item
 $\theta \in R^d$ has the isotropic centered Gaussian prior  $ \mathcal{N}( 0, \Omega^{-1} )$,  where $\Omega$  is a  $d\times d$ precision matrix. 
 \end{itemize}

The penalized EM-updates for MAPE are given by
 \begin{align}&\label{eq:sigmaMAP}
(\sigma^{(k+1)})^2 =   \biggl(  \frac 1 2 \sum\limits_{i=1}^m \bigl(   (S_0^{(k)})^2 \exp\bigl( 2 Z_i \theta^{(k)} \bigr)  + Y_i^2 \bigr) \biggr) \bigg/ \biggl(
  \sum\limits_{i=1}^m  (2 \langle N_i \rangle^{(k)} +1 ) + 1 \biggr) \;
\end{align} 
and
\begin{align}\label{eq:S0MAP}
 (S_0^{(k+1)})^2 =   \biggl( \sum\limits_{i=1}^m \langle N_i \rangle^{(k)} + c_1\biggr)
   \bigg / \biggl( \frac 1 {    2( \sigma^{(k)} )^2       }\sum\limits_{i=1}^m \bigl(\exp\bigl( 2 Z_i \theta^{(k)} \bigr) +c_2 \biggr) \; . 
\end{align}
Additionally, this gives the modified score and Fisher scoring 
 \begin{align*}
  \tilde{S}(\theta) = S(\theta) - \Omega \theta, \quad\mbox{ and }  \tilde{J} = J(\theta) + \Omega , \quad \mbox{ respectively. } 
 \end{align*}

Under our Bayesian model with weak priors the MAP estimation Eq. \eqref{eq:sigmaMAP} and Eq. \eqref{eq:S0MAP} are similar as the ML updates Eq. \eqref{eq:sigma} and Eq. \eqref{eq:S0}. Indeed, usually $\sum\limits_{i=1}^m \langle N_i \rangle \gg 1$, and we can omit the difference between Eq.(\ref{eq:sigma}) and Eq.(\ref{eq:sigmaMAP}). Then when $c_1$ and $c_2$ are small enough,
the difference between the likelihood and posterior mode of $S_0$, expressed in Eq.\eqref{eq:S0} and Eq. \eqref{eq:S0MAP} respectively, can also be ignored.
The only difference when updating $\theta$, is that we have considered the correction between the elements of a tensor represented by the prior distribution, the inverse covariance matrix, $\Omega$. Such correction may be ignorable.

Remark: Sometimes the MLE can be treated as a special case of the MAPE where the precision of the parameters depend on the chosen prior. If the effects of the priors are weak enough to be ignored, then the posterior distribution is asymptotically approximated by the likelihood. The consequence is that numerically the MAP tend to the ML estimates numerically. Such remark is not unusual (see \cite{sparacino2000}) but nearly has never appeared in the DTI literature.

\section{Results}
\label{sec:res}
\subsection{Synthetic Data}
\begin{figure}[!htb] 
\begin{minipage}[b]{1.0\linewidth}
  \centering
  \centerline{\includegraphics[width=12.5cm]{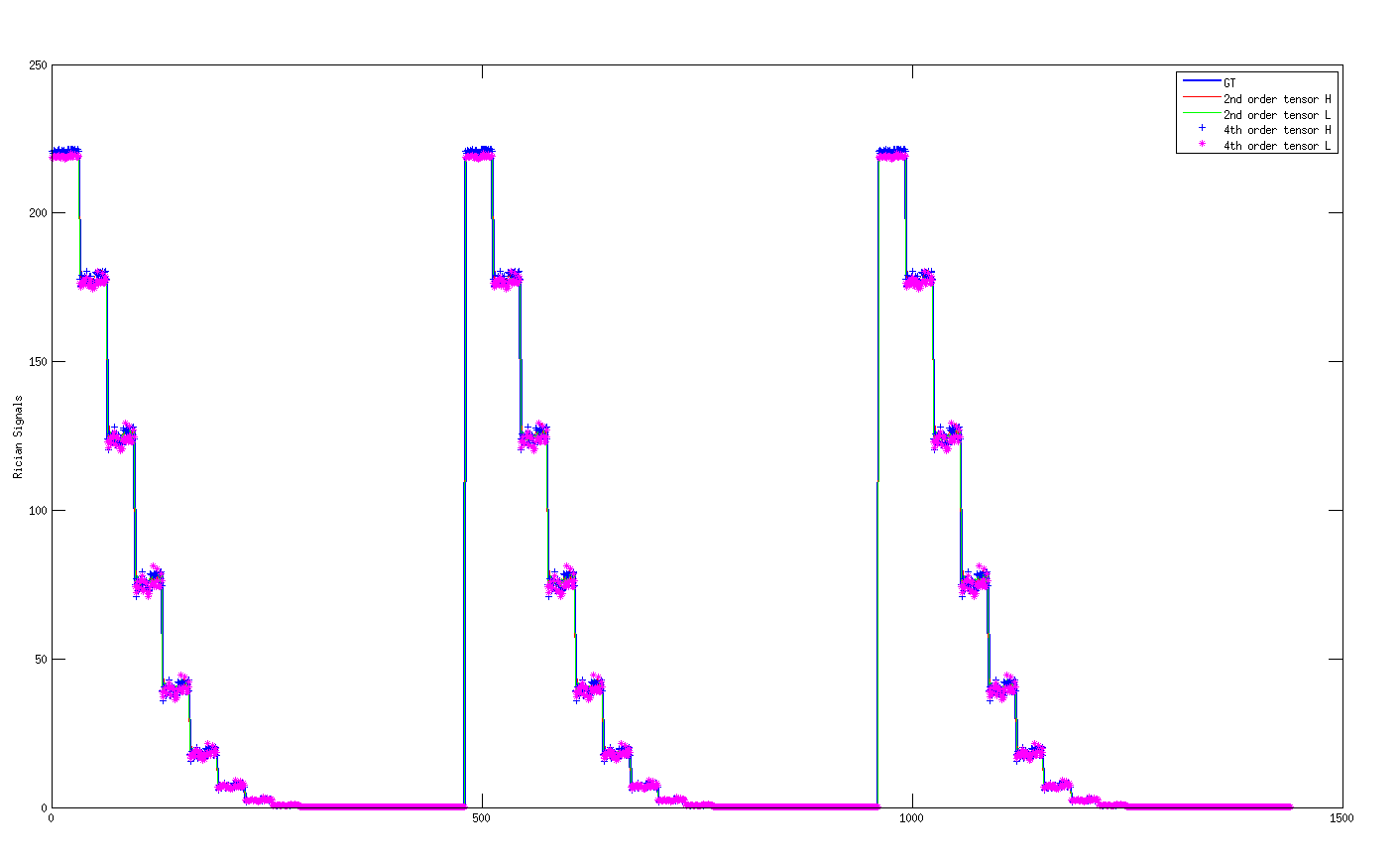}}
\end{minipage}%
\caption{Fitted Rician signals by the proposed MLE method. The blue curves depict the signal intensities of the GT. The red and green curves show the fitted signals under the 2nd-order tensor model from the high- and low- noise level datasets. Correspondingly, the black crossings and the cayn stars are the empirical values under the 4th-order tensor model.}
\label{fig:Sfit}
\end{figure}

\begin{figure}[!htb] 
  \centering
  \centerline{\includegraphics[width=8.5cm]{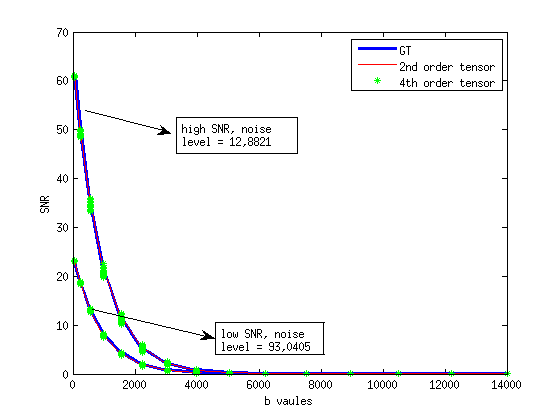}}
\caption{Empirical SNR as functions of $b$ values. The GT of the upper curve is from the high SNR corresponding the lower noise level with $\sigma^2 = 12,8821$. The bottom one has the high noise level with $\sigma^2 = 93,0405$. The red curves are fitted SNR under the 2nd-order tensor model. While the green stars represent the empirical SNR under the 4th-order tensor model.}

\label{fig:SNRwh}
\end{figure}

\begin{figure}[!htb] 
\begin{minipage}[b]{1.0\linewidth}
  \centering
  \centerline{\includegraphics[width=8.5cm]{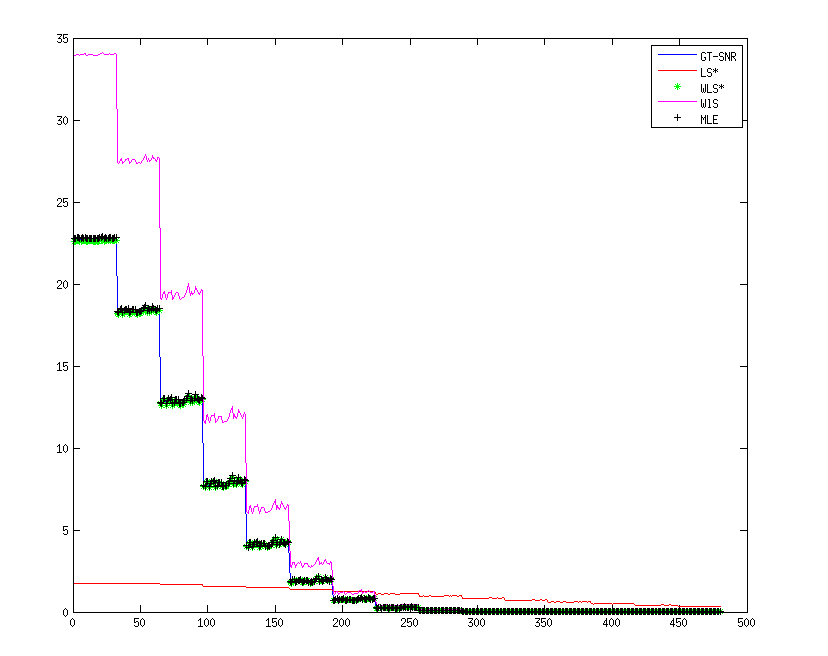}}
\end{minipage}%
\caption{Sample mean of SNR. The sample means are calculated from 100 simulated datasets. The SNR are estimated by different methods. The blue curve represents the GT. The cyan curve and the green stars are the estimates by the LS and the WLS with the truncated datasets, respectively. The red curve is the results through the WLS, and the black crossings are the empirical values by our MLE method.}
\label{fig:SNR}
\end{figure}

\begin{figure}[!htb] 
\begin{minipage}[b]{.8\linewidth}
  \centering
  \centerline{\includegraphics[width=6.5cm]{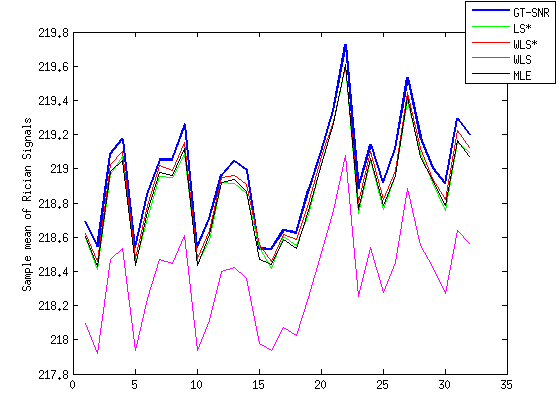}}
  \centerline{(a) low $b$ value, $b= 62  s/mm^2$}\medskip
\end{minipage}
\hfill
\begin{minipage}[b]{0.8\linewidth}
  \centering
  \centerline{\includegraphics[width=6.2cm]{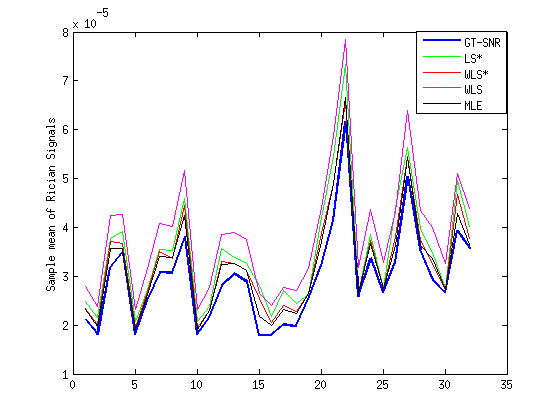}}
  \centerline{(b) high $b$ value, $b= 14000  s/mm^2$}\medskip
\end{minipage}
\caption{Sample mean of the Rician signals from low and high $b$ values. The plots illustrate the means of the signal intensities at $b= 62 ~\text{and}~ 14000 s/mm^2$, respectively, estimated by by different methods.}
\label{fig:S}
\end{figure}

\begin{figure}[!htb] 
\begin{minipage}[b]{1.0\linewidth}
  \centering
  \centerline{\includegraphics[width=8.5cm]{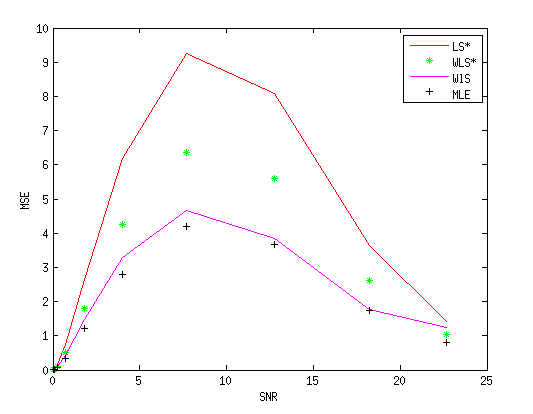}}
\end{minipage}%
\caption{MSE on tensor coefficients}
\label{fig:mse}
\end{figure}

Synthetic data sets were simulated by choosing a positive tensor of 2nd-order and of 4th-order with fixed  $S_0$ and the noise variance $\sigma^2$. The simulated data sets in the experiments arise from models with parameter values resembling the real scenario. Every dataset contains 1440 measurements which were sampled from 32 distinct gradients and 15 distinct increasing  $b$ values (knots) up to 14000$s/mm^2$, each repeated three times. The ground truth (GT) of high (H-) and low (L-) Rician noise, $\sigma^2$, are 93,0405 and 12,8821, respectively.  
To compare the performance, we first plot the ML estimated Rician signals and the GT shown in Fig. \ref{fig:Sfit}. Fig.\ref{fig:SNRwh} shows the empirical signal to noise ratio (SNR) of the distinct $b$ values from the first 480 measurements as an illustration.
For comparison of the methods, we simulate 100 datasets from high noise case under the 4th-order tensor and compare the first 480 measurements of the sample means of SNR and the GT under different methods in Fig.\ref{fig:SNR}, where 
"$*$" denotes that only the low frequencies ($b$ values less than 1000$s/mm^2$) are considered in the estimation. This figure reveals that our MLE and the WLS under a truncated dataset fit the GT well. However, when comparing the 
mean square errors (MSE) of the noise variance, the WLS* has a huge bias, 54.777, compared with the MSE of the ML estimates, 10.358. The fitted Rician signals are depicted in Fig. \ref{fig:S},
where the estimated signals are retrieved from a low b and a high b -value cases estimated from the first 480 measurements. It reveals that in the low b-value case both the WLS* and the MLE perform well.
But our ML estimates show advantages in the high b-value case. The reason is that for the WLS* we use data information which do not consider the high frequencies, but only fit back to fetch reliable estimates of signals. We further check the MSE on tensor coefficients from the 15 distinct $b$ values. The results are described in Fig.\ref{fig:mse}, where we can conclude that the MLE is the best one compared with other methods. The average computational time of the aforementioned MLE method under the 4th-order tensor model is 0.4868 seconds, which is extremely shorter than the minutes running time per voexl from the current standard methods such as MATLAB Nelder-Mead based or gradient-based estimators (see \cite{ghosh2014, landman2007}).


\subsection{Real Data}
\begin{figure}[!htb] 
\begin{minipage}[b]{1\linewidth}
  \centering
  \centerline{\includegraphics[width=9cm]{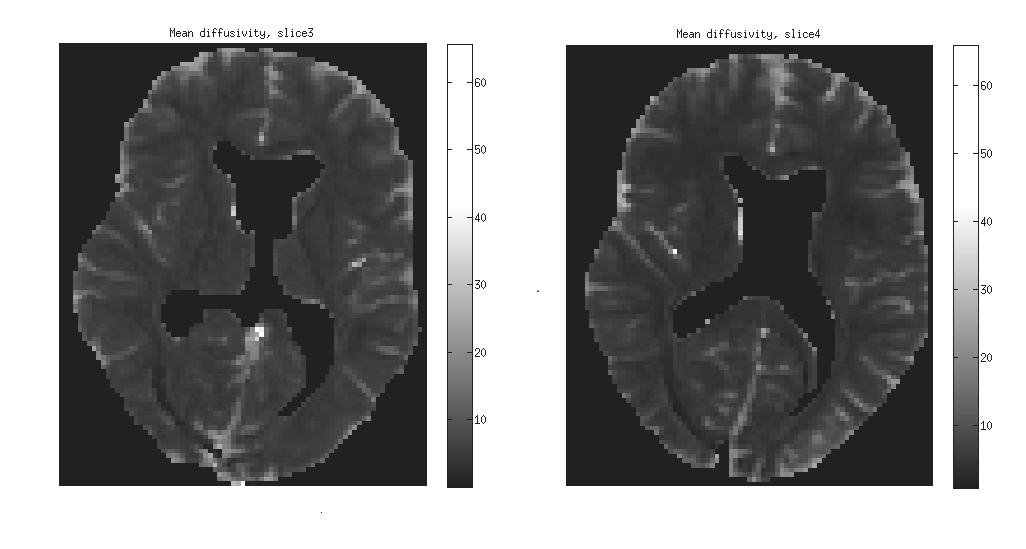}}
  \centerline{(a) Mean diffusivity (MD)}\medskip 
\end{minipage}
\hfill
\begin{minipage}[b]{1\linewidth}
  \centering
  \centerline{\includegraphics[width=8cm]{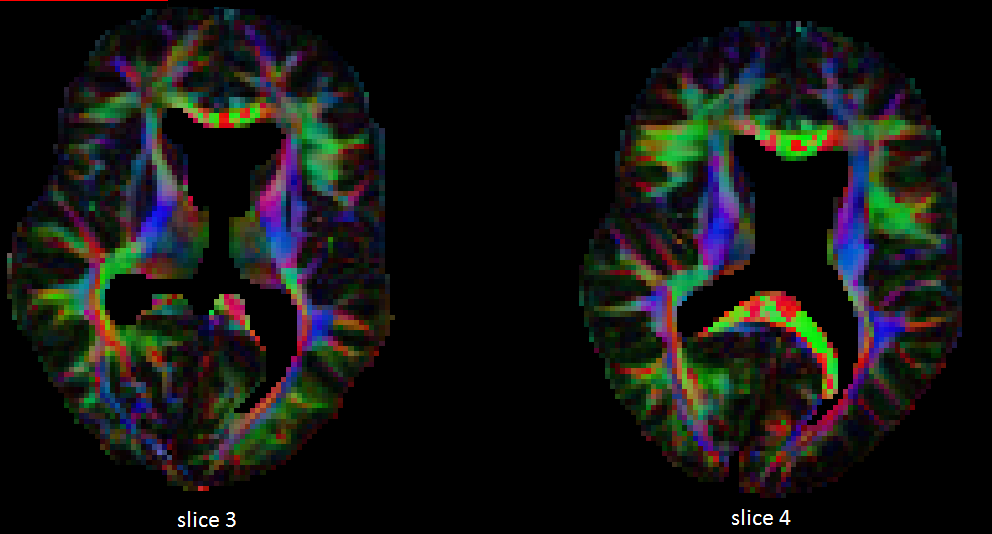}}
  \centerline{(b) Fractional anisotropy (FA)}\medskip  
\end{minipage}
\caption{MD and FA maps from two consecutive slices, where the estimated FA are computed under the 2nd-order tensor model. The color in FA represents the orientations of the fibers. The color coded FA maps are drawn by using the software ExploreDTI \cite{leemans2009}. The corresponding MD maps are from the results under the 4th-order tensor model, where the white spots corresponding to the corrupted data (artefacts) with measured magnitudes increasing at high $b$-value.}
\label{fig:FAMD}
\end{figure}

The data consist of $4596$ diffusion MR-images of the brain of an  healthy human volunteer,
 taken from four $5mm$-thick consecutive axial slices, and measured using a
 Philips Achieva $3.0$ Tesla MR-scanner.
The image resolution is
$128\times 128$ pixels of size $1.875\times 1.875$ $mm^2$. 
 After masking  out the skull and the ventricles, 
we remain with a region of interest (ROI) containing $18764$ voxels.
 In the protocol, we used all the combinations of the
$32$ gradient directions with the  $b$-values  varying in the range $0-14000 s/mm^2$, with $2-3$ repetitions, for a  total of 23\,323\,644 data points. The average computational cost per voxel by our method the 4th-order tensor model from this dataset is 1.8331 seconds.
We illustrate the results mainly under the 4th-order tensor model. Fig.\ref{fig:FAMD} shows the mean diffusivity (MD) and the fractional anisotropy (FA) of diffusion from two consecutive slices, where FA is computed from the 
results under 2nd-order tensor model, which is given by
\begin{eqnarray}\label{fa}
\text{FA} = \frac{\sqrt{3( (\lambda_1- E[\lambda])^2+(\lambda_2- E[\lambda])^2+(\lambda_3- E[\lambda])^2 )}}{\sqrt{2( \lambda_1^2+\lambda_2^2+\lambda_3^2 )}}.
\end{eqnarray}
The average values of FA from these two ROI are $0.2769 mm^2/s$ and $0.2861 mm^2/s$, respectively. The color in FA represents the orientations of the fibers. Under the 4th-order tensor model, MD is expressed as
\begin{eqnarray}&&\label{md}
\text{MD} = \frac 1 5 (D_{1111}+D_{1122}+D_{1133}+2D_{2222}+2D_{3333} \notag\\&& 
+2D_{2233})  = \frac 1 5 \text{trace} (D).
\end{eqnarray}
%
The average values of MD from Slice 3 and 4 are 6.248e-03 $mm^2/s$, 6.045e-03 $mm^2/s$, respectively, and we have the same estimated values of MD under 2nd-order tensor model. 

\begin{figure}[!htb] 
\begin{minipage}[b]{1.2\linewidth}
  \centering
  \centerline{\includegraphics[width=10cm]{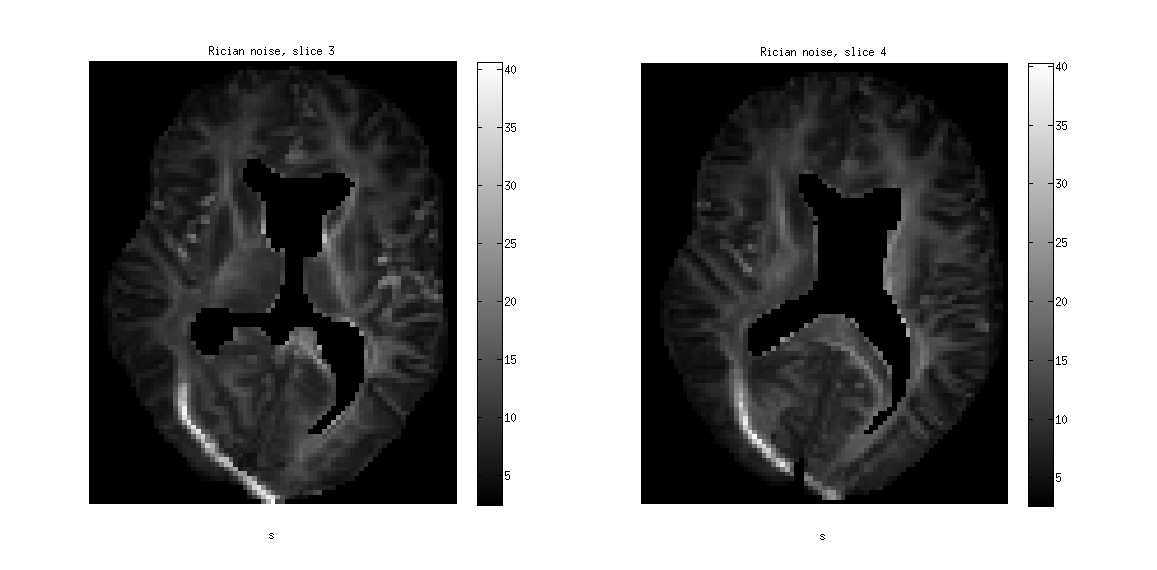}}
\end{minipage}%
\caption{Rician noise map from two consecutive slices. The white curves in the left bottom of the slices depict the artefacts corresponding to very high noise.}
\label{fig:rnoise}
\end{figure}

We also plot the Rician noise map of $\sigma$ from the two consecutive slices shown in Fig. \ref{fig:rnoise},
where the artefacts are clearly depicted by white color representing very high noise, which reveals the true scenario from the raw MR images.  

\begin{figure}[!htb] 
\begin{minipage}[b]{0.8\linewidth}
  \centering
  \centerline{\includegraphics[width=9cm]{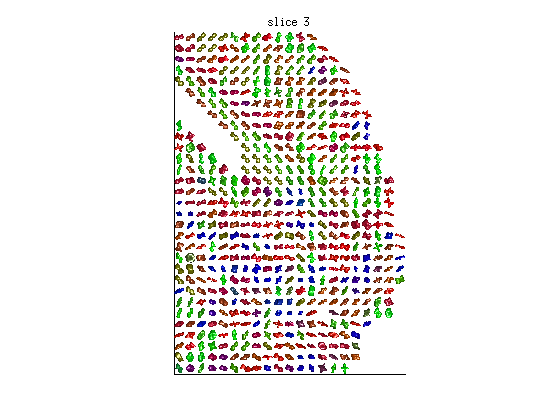}}
\end{minipage}
\hfill
\begin{minipage}[b]{0.8\linewidth}
  \centering
  \centerline{\includegraphics[width=9cm]{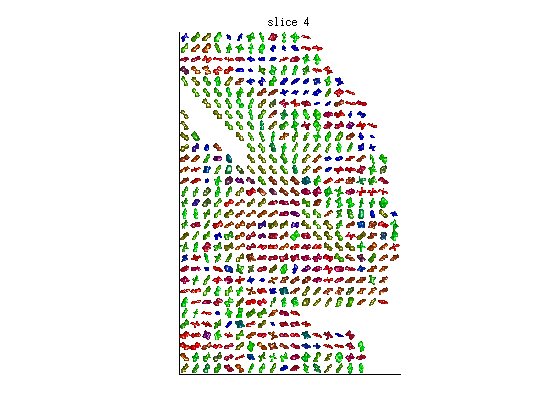}}
\end{minipage}
\caption{Visualization of DTI data with 4th-order tensors from a ROI. The color-code represents the main direction of the principal eigenvalue of the 2nd-order tensor: Red, left-right; Green, anterior-posterior; Blue, superior-inferior.}
\label{fig:tensor}
\end{figure}

Visualization of angular resolution of DTI data under different tensor models from the region of interest (ROI) of two consecutive slices are displayed in Fig. \ref{fig:tensor}, where the ROI is near the hippocampus and the empty spaces inside of left parts of the diffusion profiles (DP) are the masked ventricle. DP depict under the 4th-order tensors providing much angular information of diffusion, where the colors represents the principle orientations of diffusion at each voxel. These tensor profiles are plotted by MATLAB fanDTasia toolbox \cite{barmpoutis2007}.

\section{Discussion}
\label{sec:dis}
Our method substantially differs from the previous ones in the literature and the advantages are summarized by the following points:
1) We introduce a novel data augmentation, which allows the non-linear regression problem to be transformed into the GLM framework in DTE. 2) Subsequently, the computation is dramatically reduced due to the tractable modes of parameters of interest in the sense of point estimation. In addition, when employing Fisher-scoring scheme we simplify the complexity of the Fisher information. 3) Our Rician noise model can be combined with any tensor model in different representation, such as spheric harmonic expansion, by reparametrization. 4) Either ML or MAP estimation yields more accurate estimates than the LS and WLS do. In addition, high frequencies from the low SNR data and the zero measurements are also included into the estimation. These data are known to contain detailed anatomical information of the complex tissue in vivo. 5) Our method leads to significantly less biased estimates of the noise level, which plays key role in denoising the MRI and cleaning the artefacts.

{\it Positive Constraints.} The physical feature of diffusion requires the tensor to be positive definite.
Our model allows to check the positivity of diffusivity in the tensor updates 
under the scheme of Fisher-scoring method. For the rank-2 tensor model, the constraining is fairly easy 
to do by computing the eigenvalues of the tensor matrix $D$. For HARDI, Barmpoutis et al. \cite{barmpoutis2009}
propose the Gram matrix approach, using the quartic form to guarantee the positivity.
Other methods such as \cite{qi2010higher} address the constraint by calculating the Z-eigenvalue polynomials.  

{\it MLE VS MAPE}. In this work, we did not list the results from MAPE but we emphasize
the differences between these two methods. Bayesian methods have advantages in the learning process, 
meaning that they may gain extra information from the prior knowledge. When the prior is weak, like in our case, we learn things from the data, what we actually do when approaching the problem through frequentist statistical modeling. In order to learn the uncertainty of the diffusion parameters, a fully Bayesian approach is highly recommended to characterize the posterior parameter distributions rather than point estimation.

\section{Acknowledgement}
We thank Professor Antti Penttinen for reviewing the manuscript and providing insightful comments. We would also like to thank the Radiology Unit of Helsinki University Hospital for the data collection. This work was funded by Doctoral Program in Computing and Mathematical Sciences (COMAS), University of Jyv\a"skyl\"a. We acknowledge the Finnish Doctoral Programme in Stochastics and Statistics (FDPSS) provided travel funds for this research.

\appendix
\renewcommand{\thesubsection}{\Alph{subsection}}
\numberwithin{equation}{subsection}
\renewcommand{\theequation}{\Alph{subsection}.\arabic{equation}}
   
\section*{Appendix}
\subsection{Theory of the EM algorithm}\label{ape:AA}
Consider a statistical model $\bigr(p_{\theta}(y), \theta \in \Theta\bigl)$, where $\Theta \subseteq \R^d$, and
the likelihood of the observed data $y=(y_1,\dots,y_n)$ is expressed as the marginal of an integrated joint likelihood
\begin{align*}
p_{\theta}( y) = \int_{ {\mathcal Z} } p_{\theta}(z,y) dz \; .
\end{align*}
Here $z=(z_1,\dots,z_n)\in {\mathcal Z}$ and  $z_i$ are interpreted as latent  variables. When ${\mathcal Z}$ is discrete, we replace integrals by  sums.
In the EM algorithm \cite{dempster}, starting with an inital value $\theta^{(0)}\in \Theta$, we iterate the maximization step
\begin{align} \label{EM:iteration}
   \theta^{(k+1)} = \arg\max_{\theta \in \Theta} 
   \biggl\{  
   E_{\theta^{(k)} }\bigl( 
   \log p_{\theta}( z,y )  
   \big\vert y \bigr)
   \biggr\} = \arg\max\limits_{\theta \in \Theta} 
   \biggl\{  \int\limits_{ {\mathcal Z} }   \log p_{\theta}( z,y )  p_{\theta^{(k)}}( z | y) dz   \biggr\}\; , 
\end{align}
where the integration is  with respect to the conditional density 
\begin{align*}
p_{\theta^{(k)}}( z | y)=\frac{ p_{\theta^{(k)}}(z,y) }
{ p_{\theta^{(k)}}(y) } \quad \mbox{ (Bayes formula). } 
\end{align*}
By Jensen inequality, 
the Kullback relative entropy of the conditional distribution $ p_{\theta}( z | y )$ related
 to  $p_{\theta^{(k)} }( z | y )$, given by
\begin{align*}&  
K( \theta^{(k)},\theta|y ):=
 E_{\theta^{(k)} }
 \biggl(\log \biggl( 
 \frac{p_{\theta^{(k)}} ( z | y ) }{ 
 p_{\theta}( z | y )}  \biggr)  \bigg\vert y  \biggr)=  
 \int_{{\mathcal Z} } \log\biggl(  \frac {p_{\theta^{(k)}}( z | y ) } { p_{\theta}( z | y ) } \biggr) p_{\theta^{(k)}}(z|y) dz  \; ,
 \end{align*}
 is non-negative, 
 which implies  
\begin{align}\label{eq:optimizing} & 
 \log p_{\theta}(y) -\log p_{\theta^{(k)}}(y)  \ge 
  & \nonumber \\ &  
  \int_{{\mathcal Z} } \log\bigl( p_{\theta}( z , y )  \bigr) p_{\theta^{(k)}}(z|y) dx -   
 \int_{{\mathcal Z} } \log\bigl( p_{\theta^{(k)}}( z , y ) \bigr) p_{\theta^{(k)}}(z|y) dx \; ,
&\end{align}
and consequently
\begin{align*}
 \log p_{\theta^{(k+1)}}(y) \ge \log p_{\theta^{(k)}}(y)  \;,   
\end{align*}
i.e. the  EM-step does not decrease the marginal likelihood of $y$.
It follows also  from  \eqref{eq:optimizing}, that  fixing a  $\theta$-subvector  and maximizing with respect to
the remaining $\theta$-coordinates does not decrease the marginal likelihood of $y$.
The EM algorithm is iterated until convergence to a fixed point $\theta^{(\infty)}$,
a local maximum of the marginal likelihood $p_{\theta}(y)$. When the local maximum
is the global one, $\widehat\theta_{ML}= \theta^{(\infty)}$ is the maximum likelihood estimator of the parameter.
The advantage of the EM algorithm is that,
for some smart choices of the data augmentation $z$  and the joint density $p_{\theta}(z,y)$, 
the maximization  step \eqref{EM:iteration} can be simpler than maximizing directly the  marginal likelihood $p_{\theta}(y)$,
especially in cases where the latter is hard to evaluate.

\subsection{MLE by the EM algorithm in DTI}\label{ape:A}
Appendix \ref{ape:A} gives details of the expectation-maximization (EM) algorithm in DTE. 
We consider the Rician noise model with the Poissonian data augmentation of Section \ref{sec:theo}. The latent augmented variable $N$ conditionally on $X,Z$ is given by
\begin{align*}
  p_{t,\sigma} ( N=n |X ,Z )  =\frac 1 { I_0(2\tau) } \frac{Â \exp(-2 \tau) \tau^{2n} }{ (n!)^2 } ,  \, n \in \N, \mbox{ with } 
  \quad\tau= X \sqrt{ \frac{  t }{2\sigma^2} }\;
\mbox{ and } \quad  X= Y^2\; .
\end{align*}
It follows \cite{Gasbarra} that this discrete distribution is referred as reinforced Poisson distribution with parameter $\tau$.

In the EM algorithm we need to compute the conditional expectation of $N$ conditionally on $X$ and the design matrix $Z$. 
Given the current values $t^{(k)},~{\sigma^2}^{(k)}$ 
\begin{eqnarray*}&&
\langle N \rangle^{(k)} :=  E_{t^{(k)},{\sigma^2}^{(k)}}\bigl( N\big\vert  X, Z \bigr) 
 = \sum_{n=1}^{\infty} n  p_{t,\sigma} ( N=n | X, Z )\\&& 
=\tau^{(k)}/ 2 \frac{d}{d \tau^{(k)}}\log {}_0F_1(1, (\tau^{(k)})^2) 
  = \tau^{(k)}/ 2 \frac{d}{d \tau^{(k)}}\log J_0(2 \tau^{(k)} \sqrt{-1} )
= \frac{ \tau^{(k)} J_{-1}(  2 \tau^{(k)} \sqrt{-1} )}{    J_{0}(  2\tau^{(k)} \sqrt{-1} )}\\&&  
 = \frac{ \tau^{(k)}  \; I_1( 2 \tau^{(k)} ) }{ I_0( 2\tau^{(k)})} \; ,
\end{eqnarray*}
with
\begin{eqnarray*}
t^{(k)} = t({S_0^2}^{(k)}, \theta^{(k)},{\sigma^2}^{(k)})= \frac{  {S_0^2}^{(k)} \exp(2 Z\theta^{(k)})  }{ 2{\sigma^2}^{(k)} }  \mbox{ and } \quad \tau^{(k)} =\frac{\sqrt X_i }{ {\sigma^2}^{(k)} \sqrt{2} } \exp( Z_i \theta^{(k)} ){S_0}^{(k)}.
\end{eqnarray*}
Note that ${}_0F_1(1,\tau^2) = J_0(2\tau\sqrt{-1})=I_0(2\tau)$, where $J_0(z)$ is the zero-order Bessel function of the first kind, $I_0(z)$ is the zero-order modified Bessel function of first kind, which satisfies 
\begin{eqnarray*}
J_v'(x)  = J_{v-1}(x)-\frac {\nu}{x}J_v(x),
\end{eqnarray*}
and
\begin{eqnarray*}
J_{-n}(x)=(-1)^n J_{n}(x), \qquad  I_n(z)=i^{-n} J_n(z i ).
\end{eqnarray*}
 
In the M-step, we maximize the parameters of the augmented log likelihood $Q$ from Eq. \eqref{eq:joint} w.r.t $(\theta,\sigma^2, S_0^2)$. Omitting the items not depending on these parameters, $Q$ can be expressed as
\begin{align} \label{loglik}
 \sum_{i=1}^m
\biggl(  \log(S_0^2) -2\log(\sigma^2) + 2 Z_i\theta \biggr) \langle N_i \rangle^{(k)} -m \log(\sigma^2)
-  \frac 1{ 2 \sigma^2}  \sum_{i=1}^m \bigl(S_0^2\exp(2Z_i\theta) + X_i \bigr)  \; . 
\end{align} 
It is easy to see in Eq. \eqref{loglik} that the log likelihood w.r.t $\sigma^2$ and $ S_0^2$ are inverse Gamma and Gamma distributions, respectively. Hence, we update these two parameters by their modes:
\begin{align}&\label{score_sigma}
\widehat{\sigma^2}_{ML} := \arg{\max_{\sigma_g^2}(Q)} =
\frac{ \sum_{i=1}^m (X_i + \exp(2 \hat{\theta} Z_i )\hat{S_0}^2 )} {2\sum_{i=1}^m (2\langle N_i \rangle +1)}
\end{align}
and
\begin{align}\label{score_S02}
 \widehat{S_0^2}_{ML} := \arg{\max_{S_0^2}(Q)} = \frac{2\widehat{\sigma^2}_{ML} \sum_{i=1}^m \langle N_i \rangle}{\sum_{i=1}^m \exp(2 Z_i \hat{\theta})}. 
\end{align}

To apply the Fisher scoring method, we have the score of $\theta$ is
\begin{eqnarray}&& \label{score_theta}
S(\theta)=2 \sum_{i=1}^m\langle N_i \rangle Z_i - \frac{\widehat{S_0^2}_{ML}}{\widehat{\sigma^2}_{ML} } \sum_{i=1}^m \exp( 2 Z_i \theta) Z_i,
\end{eqnarray}  
\medskip
and the Fisher-information is given by
\begin{eqnarray}&& \label{fisher_theta}
J(\theta) = E \biggl[ - \frac{\partial^2 Q}{\partial \theta_h \partial \theta_k }\biggr] =  \frac{\widehat{S_0^2}_{ML}}{\widehat{\sigma^2}_{ML} } \sum_{i=1}^m \exp( 2 Z_i \theta) Z_i Z_i^T.
\end{eqnarray}  

\subsection{Maximization of Rician Log-likelihood}\label{ape:B} 
Without data agumentation, we have to directly maximize the Rician log likelihood $Q_{Rician}$, in short $Q_r$ thereafter, by using some typical MLE method, such as gradient descent. 
Then the first (the score) and second derivatives of $Q_r$ are usually required. The loglikelihood $Q_r$ is
 \begin{align*}  &
Q_r = \mbox{const.} - m\log(\sigma^2) - \frac 1 {2\sigma^2}  \sum_{i=1}^m \biggl( Y_i^2  + \exp( 2 Z_i \theta )S_0^2 \biggr)
+ \sum_{i=1}^m \log I_0\biggl( \frac{ {  Y_i } \exp( Z_i \theta)\sqrt{S_0^2}   }{\sigma^2} \biggr),
& \end{align*}
where  $I_k( \tau)$ are modified Bessel functions of first kind satisfying 
$$
I_0^{'}(\tau)=I_1(\tau), \qquad I_0^{''}(\tau)=I_1^{'}(\tau) = ( I_0(\tau)+I_2(\tau))/2.
$$

\bigskip
The score of $\sigma^2$ and $S_0^2$ are respectively given by
\begin{align} &\label{score_sigmar}
\frac{ \partial Q_r}{\partial \sigma^2} = - \frac m {\sigma^2}  + \frac 1 {2\sigma^4} 
\sum_{i=1}^m \biggl( Y_i^2 + \exp( 2 Z_i \theta_i )S_0^2  \biggr)  
-\frac 1 {\sigma^4}  \sum_{i=1}^m  g \biggl(   Y_i \exp( Z_i \theta)S_0  \sigma^{-2} \biggr)   Y_i \exp(Z_i \theta)S_0  
\end{align}
and 
\begin{align}& \label{score_S0r}
\frac{ \partial Q_r}{\partial S_0^2} = -  \frac 1 {\sigma^2} 
\sum_{i=1}^m \exp( 2 Z_i \theta_i ) 
 + \frac 1 {2 \sigma^2 \sqrt{S_0^2}}  \sum_{i=1}^m  g\biggl(  Y_i  \exp( Z_i \theta)S_0 \sigma^{-2} \biggr)   Y_i \exp( Z_i \theta). 
\end{align} 
The score of $\theta$ is given by
 \begin{align} & \label{score_thetar}
\frac{ \partial Q_r}{\partial \theta_k} = -  \frac {S_0^2} {\sigma^2}
\sum_{i=1}^m \exp( 2 Z_i \theta_i )  Z_{ik} 
 + \frac 1 {\sigma^2}  \sum_{i=1}^m  g \biggl(   Y_i \exp( Z_i \theta)S_0  \sigma^{-2} \biggr)   Y_i \exp(Z_i \theta)S_0 Z_{ik}. 
\end{align} 
The Hessian of $\theta$ is given by
\begin{align*} &
\frac{ \partial Q_r^2}{\partial \theta_h\partial \theta_k} = -  \frac{2 S_0^2}{\sigma^2}
\sum_{i=1}^m  \exp( 2 Z_i \theta_i )   Z_{ih} Z_{ik} 
 + \frac {S_0} {\sigma^2}  \sum_{i=1}^m   Y_i \exp( Z_i \theta)  Z_{ik} Z_{ih} \biggl\{  g\biggl(  Y_i  \exp( Z_i \theta)S_0 \sigma^{-2} \biggr) 
 & \\ & + g^{'}\biggl(   Y_i  \exp( Z_i \theta) S_0 \sigma^{-2} \biggr) \frac{ Y_i \exp( Z_i \theta) S_0 }{\sigma^2} \biggr\} 
& \\& = \sum_{i=1}^m Z_{ih} Z_{ik}  \biggl( -4 t_i^2 + \tau_i (g(\tau_i) + \tau_i g^{'}(\tau_i) \biggr)
 = \sum_{i=1}^m Z_{ih} Z_{ik}  \biggl( -4 t_i^2   + \tau_i^2 - \tau_i^2 \biggl( \frac{I_1(\tau_i)}{I_0(\tau_i)}\biggr)^2 \biggr).
\end{align*}
where  we denote 
\begin{align*} &
\tau_i = \frac{Y_i \exp(Z_i \theta)S_0}{2\sigma^2}, \qquad
g(\tau) =  \frac{ d}{ d\tau} \log I_0(\tau)=\frac{ I_1(\tau)  }{ I_0(\tau) },
&  \\ &
g^{'}(\tau)= \frac{ d^2}{ d\tau^2} \log I_0(\tau)  = \frac 1 2 \biggl( 1+ \frac{ I_2(\tau)  }{ I_0(\tau) } \biggr)
 - \biggl(\frac{ I_1(\tau)  }{ I_0(\tau) }  \biggr)^2 
= 1 - \frac{ I_1(\tau)  }{ I_0(\tau) }   - \biggl(\frac{ I_1(\tau)  }{ I_0(\tau) }  \biggr)^2 
&  \\ &
\text{with}
&  \\ &
I_2(\tau) =  I_0(\tau)-\frac{2 I_1(\tau)}{\tau}.
\end{align*}
For  SNR $> 10$, the corresponding Fisher-information matrix is  
approximated   by
\begin{eqnarray} \label{fisher_rice}
I_r(\theta) =& E \biggl[- \frac{ \partial Q_r^2}{\partial \theta_h\partial \theta_k} \biggr]
 \approx \sum\limits_{i=1}^m Z_{ih} Z_{ik} \biggl( \frac{S_0^2}{\sigma^2} \exp(2 Z_i \theta) - \frac 1 2 \biggr),
\end{eqnarray}
where 
(see\cite{andersson2008})
\begin{align*}&  
E\biggl[ \tau_i^2 \biggl( \frac{ I_1(\tau_i)  }{ I_0(\tau_i) }  \biggr)^2  \biggr] \approx \biggl(\frac{S_0^2}{\sigma^2} \exp(2 Z_i \theta) \biggr)^2 
+\frac{S_0^2}{\sigma^2} \exp(2 Z_i \theta) - \frac 1 2 .
\end{align*}

\subsection{Method Comparison}
In this section, we discuss the differences between our data-augmentation based on the EM algorithm and on the typical MLE method through direct maximization at $Q_r$.
\begin{enumerate}
 \item We do not need to calculate all the elements of the Hessian as we can directly find the modes of $S_0^2$ and $\sigma^2$ by data augmentation. A small improvement appears in the
reparametrization of  $S_0$ or $\log S_0$ by $S_0^2$.
\item In the E-step we compute 
\begin{eqnarray}\label{expections}
\langle N_i \rangle =
E_{\theta^{(k)}, {\sigma^2}^{(k)},{S_0^2}^{(k)}} ( N_i| Y_i), 
\end{eqnarray}
which does not depend on the parameters $\theta,\sigma^2$ and $S_0^2$. In the M-step we use Eq. \ref{expections},
the recursive values from $\theta^{(k)}, {\sigma^2}^{(k)},{S_0^2}^{(k)}$, instead of solving the intractable formula w.r.t those parameters. That dramatically reduces the computation of the score from Eq.(\ref{score_S0r},\ref{score_sigmar},\ref{score_thetar}) to Eq.(\ref{score_S02},\ref{score_sigma},\ref{score_theta}), respectively.
\item The
EM algorithm allows us to use empirical values from Eq.\eqref{expections} to compute the Fisher information. Our Fisher information $J(\theta)$ which fits the whole range of SNR and is slightly bigger than the approximated one, $I_r(\theta)$,
expressed in (Eq. \eqref{fisher_rice}), which requires heavy mathematical calculations to deal with different expectations (see \cite{andersson2008} for more details). In addition, when computing the score of $\theta$ in Eq. \ref{eq:theta}, we do not need to update the items containing $N_i$ as they are fixed values from Eq.\eqref{expections}. All those lead reduced computation in practice.
\end{enumerate}


\end{document}